\documentclass[english,twocolumn,pra,eqsecnum]{revtex4}
\usepackage[T1]{fontenc}
\usepackage[latin9]{inputenc}
\usepackage{float}
\usepackage{color}
\usepackage{graphicx}
\usepackage{amssymb}

\usepackage{babel}

\begin{document}

\title{Dynamical Quantum Memories}

\author{Q. Y. \surname{He}$^{1}$}

\author{M. D. \surname{Reid}$^{1}$}

\author{E. \surname{Giacobino}$^{2}$}

\author{J. \surname{Cviklinski}$^{2}$}

\author{P. D \surname{Drummond}$^{1,*}$}

\affiliation{$^{1}$ARC Centre of Excellence for Quantum-Atom Optics, The University
of Queensland, Brisbane QLD 4072, Australia }

\affiliation{$^{2}$Laboratoire Kastler Brossel, Universit$\acute{e}$ Paris 6,
Ecole Normale Sup$\acute{e}$rieure et CNRS, UPMC Case 74, 4 place
Jussieu, 75252, Paris Cedex 05, France}

\affiliation{$^{*}$Email: drummond@physics.uq.edu.au}

\begin{abstract}
We propose a dynamical approach to quantum memories using an oscillator-cavity
model. This overcomes the known difficulties of achieving high quantum
input-output fidelity with storage times long compared to the input
signal duration. We use a generic model of the memory response, which
is applicable to any linear storage medium ranging from a superconducting
device to an atomic medium. The temporal switching or gating of the
device may either be through a control field changing the coupling,
or through a variable detuning approach, as in more recent quantum
memory experiments. An exact calculation of the temporal memory response
to an external input is carried out. This shows that there is a mode-matching
criterion which determines the optimum input and output mode shape.
This optimum pulse shape can be modified by changing the gate characteristics.
In addition, there is a critical coupling between the atoms and the
cavity that allows high fidelity in the presence of long storage times.
The quantum fidelity is calculated both for the coherent state protocol,
and for a completely arbitrary input state with a bounded total photon
number. We show how a dynamical quantum memory can surpass the relevant
classical memory bound, while retaining a relatively long storage
time. 
\end{abstract}
\maketitle
Quantum memories are devices that can capture, store, then replay
a quantum state on demand\cite{duanlukinciraczol}. In principle,
storage is not a problem for time-scales even as long as seconds or
more, since there are atomic transitions with very long lifetimes
that could be used to store quantum states\cite{Hollberg,Ye}. A
quantum memory can store quantum superpositions. These cannot be stored
in a classical memory in which a measurement is made on a quantum
state prior to storage. The fundamental interest of this type of device
is that one can decide at any time to read out the state and perform
a measurement. In this way, the collapse of a wavepacket is able to
be indefinitely delayed, allowing new tests of decoherence in quantum
mechanics.

Such devices also have a fascinating potential for extending the reach
of quantum technologies. Here, the main interest is in converting
a photonic traveling-wave state - useful in communication - to a static
form. Although atomic transitions are normally considered, actually
any type of static mode can be used as a quantum memory. For the implementation
of quantum networks, quantum cryptography and quantum computing, it
is essential to have efficient, long-lived quantum memories\cite{duanlukinciraczol}.
These should be able to output the relevant state on demand at a much
later time, with a high fidelity over a required set of input states.
The benchmark for a quantum memory is that the average fidelity $\bar{F}$
must be higher than any possible classical memory when averaged over
the input states: $\bar{F}>\bar{F}_{C}$.

The vital task of a quantum memory is to efficiently store quantum
states in a static quantum system and then retrieve them in the form
of a \emph{propagating} quantum signal - typically a photonic pulsed
field. It is also important that the read-in and read-out are in well-defined
temporal modes that are synchronized to a clock pulse. This is essential
if the stored quantum field is to be used in any further quantum logic
operations. In establishing fidelity, it is therefore necessary to
use a synchronized local oscillator measurement to determine which
temporal mode is occupied reproducibly. Essential to the principle
of the quantum memory and its role in quantum repeaters and cryptography
is that the memory is able to be read out \emph{long after the destruction
of the input state}. This leads to a second essential criterion, which
is that the memory time $T$ must be longer than the duration $T_{I}$
of the input signal: $T>T_{I}$.

The transfer of quantum information from light to atoms was demonstrated
using off-resonant interactions with spin polarized atomic ensembles\cite{Julsgaard}.
The transfer and retrieval of classical pulses\cite{claspulsememory},
photon states\cite{Chaneliere,Eisaman,Chou} and, more recently, squeezed
states\cite{Appel,honda} has been realized using atomic three-level
transitions and electromagnetically induced transparency (EIT)\cite{Fllukin}.
Promising are memories based on controlled reversible inhomogeneous
broadening (CRIB)\cite{inhomo}. Other recent experiments report improved
efficiencies\cite{hetet} using two level atoms Stark shifted by an
external electric control field. Another device type is the quantum
circuit based on superconducting transmission lines and squids, in
which the device characteristics can be fabricated as an integrated
circuit\cite{solidstcircuits,nori,lehcol}. Nanomechanical oscillator
storage is also not impossible\cite{Harris2008}, allowing the potential
for storage and retrieval of quantum superposition states in tests
of macroscopic quantum mechanics\cite{BoseJacobsKnight}.

Current experiments are frequently limited by the problem that storage
times $T$ achieving high fidelity are shorter than the time $T_{I}$
taken to capture the incoming quantum information. On the other hand,
the use of long storage times leads to rapid degradation in the retrieval
efficiency, hence giving a low quantum fidelity. A common approach
has been to consider a broadband continuous-time input. Alternatively,
where pulses have been used, input - output efficiencies are often
measured in a regime of minimal storage time, so that the memory acts
to delay, rather than store, a pulse. This problem was recognised
by Appel et al\cite{Appel}{, who report fidelities with a relative
storage time $T/T_{I}$ of order $1.6$. It} is an outstanding challenge
to design a practical quantum memory which can retain an arbitrary
quantum state with good fidelity, for on-demand synchronous readout
over times long compared to the input signal duration.

\begin{figure}
\includegraphics[width=0.9\columnwidth]{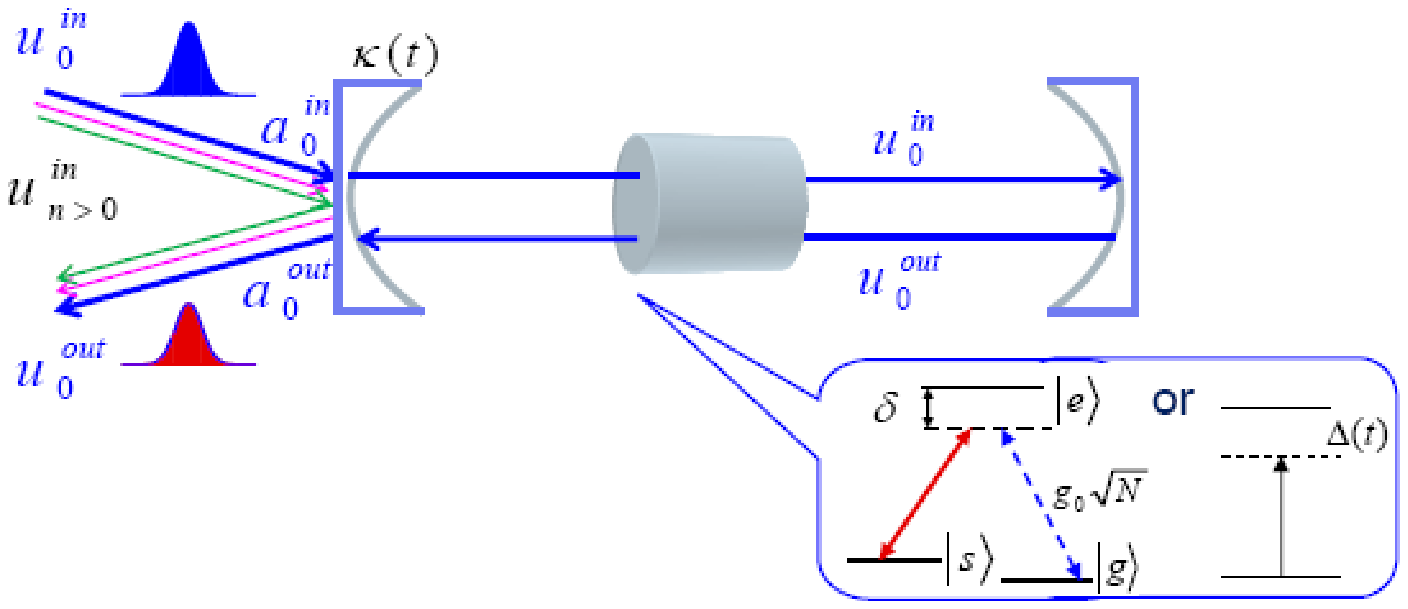}

\caption{Proposed dynamical atom-cavity memory scheme. The cavity couples effectively
to only one external incoming and outgoing mode, labelled here as
$u_{0}^{in}$ and $u_{0}^{out}$ respectively. This implies an optimal
pulse shape necessary for efficient imprinting and retrieval of the
quantum information, as represented by the mode $a_{0}$, onto and
from the atomic medium internal to the cavity. Storage is achieved
through modulation of the atom-cavity coupling $g$ or detuning $\Delta$.
\label{fig:model}}

\end{figure}

In this paper, we propose that these limitations may be overcome with
the employment of a dynamical oscillator-cavity quantum memory. While
useful in generation of squeezed and entangled states, most experimental
quantum memories have not so far focused on intra-cavity interactions\cite{storing cavity1,storing cav 2,storecav3,theorycavity2,cavityqcomp,cavpromise},
apart from recent single-photon experiments\cite{Vuletic2007,Rempe2007,Vogel2008}.
A limiting factor has been the lack of a full theoretical treatment
of the interplay between the storage medium, the cavity and the incoming
mode dynamics, together with their effect on memory performance.

Here, we bridge this gap by analyzing the memory \emph{dynamics} of
models of quantum memories, to calculate directly the memory response
in the time domain. This allows further insight over previous treatments,
which have been restricted by the assumption of slowly varying incoming
signals in an adiabatic approximation\cite{theory cavity,theory cavity 3,lulinprl2000,scottkimble,theorycavityDanpulse}.
Our theoretical approach is carried out with simple non-saturating
linear oscillator models that are analytically soluble. This strategy
can be applied to more general models, which behave as simple oscillators
for low input signal intensities.

Our conclusion is that for quantum memories employing a coupled oscillator-cavity
strategy, there is a \emph{critical} coupling between the oscillator
and cavity that gives an optimal temporal mode structure to allow
for high efficiency and fidelity of input and output states. For low
loss oscillator memories, this allows both high fidelity and long
storage times in the cavity relative to the input pulse-width. The
critical cavity coupling is closely related to the critical damping
of a harmonic oscillator. We show that one can achieve the memory
by either a modulation of the coupling or the detuning of the oscillator
mode that stores the quantum state.

For a step-function gate the corresponding temporal mode has an asymmetric
shape with duration of the order of the cavity ring-down time, which
can be fast compared to the atomic decay time. In our treatment, the
output mode is a time-reversed copy of the input. This time-reversal
of an asymmetric mode could cause problems, for example, in local
oscillator measurements or using cascaded devices. However, in a future
paper, we show that the mode-shape can be further optimized with a
time-dependent coupling, which leads to a fully time-symmetric mode
in which both the input and output modes are identical.

Our results are applicable to any technologies employing cavity-like
storage with a linear intra-cavity response. One example of this,
as indicated above, would be the case of an ensemble of atoms with
two or three-level transitions, as typically utilized in current experiments.
Other possibilities include memories using superconducting cavities
with Josephson junction qubit storage\cite{solidstcircuits}, and
states encoded into positions of atoms\cite{scottkimble}, molecules
or even nano-oscillators\cite{Vahala,MarquardtCooling,Harris2008}.
The theoretical approach developed in this paper can also be extended
to apply to spatial mode-structures\cite{Hak84,Drummond1981}, as
will be analyzed elsewhere.

\section{Linear Memory}

The quantum memory device we consider is that of a propagating single
transverse-mode field $A^{in}(t)$ entering a cavity with an atomic
or other oscillator medium (Fig. \ref{fig:model}). Writing into the
memory occurs up to a time $t=0$, during which time there is a nonzero
interaction, between field and cavity, to allow the transfer of information.
After a controllable storage time $T$, when the interaction is off,
the interaction is switched on again, so the memory reads out into
an outgoing quantum field $A^{out}(t)$ at $t>T$ (Fig. \ref{fig:Memory-involves-three}).
The present paper focuses on fields with single transverse modes that
are spatially mode-matched to the memory device\cite{Drummond1981,DruRay91}.
We consider linear memories which are agnostic with regard to the
quantum state or protocol, apart from a physical upper bound to the
pulse energy.

\subsection{Atomic example}

There are many possible implementations in which a quantum system
is coupled to an interferometer mode. To illustrate this, we first
consider the classic case\cite{Hak84,Drummond1981} of a two-level
near-resonant atomic medium, with a microscopic Hamiltonian of form:\begin{equation}
\hat{H}_{af}=\sum_{j}\hat{H}_{j}\,,\end{equation}
 where the Hamiltonian terms are given by:

\begin{eqnarray}
\hat{H}_{1} & = & \hbar\sum_{k}\omega_{k}\hat{a}_{k}^{\dagger}\hat{a}_{k}\nonumber \\
\hat{H}_{2} & = & \frac{\hbar}{2}\sum_{\mu}\omega_{\mu}(t)\hat{\sigma}_{\mu}^{z}\nonumber \\
\hat{H}_{3} & = & \hat{H}^{a}+\hat{H}^{f}\nonumber \\
\hat{H}_{4} & = & \hbar\sum_{k}\sum_{\mu}\left(g_{k}\left(t,\mathbf{r}_{\mu}\right)\hat{a}_{k}^{\dagger}\hat{\sigma}_{\mu}^{-}+H.c.\right)\nonumber \\
\hat{H}_{5} & = & \hbar\sum_{\mu}\left(\hat{\Gamma}_{\mu}^{\sigma\dagger}\hat{\sigma}_{\mu}^{-}+\hat{\Gamma}_{\mu}^{\sigma}\hat{\sigma}_{\mu}^{+}+\hat{\Gamma}_{\mu}^{z}\hat{\sigma}_{\mu}^{z}\right)\nonumber \\
\hat{H}_{6} & = & i\hbar\sum_{k}\left(\hat{\Gamma}_{k}^{a}\hat{a}_{k}^{\dagger}-\hat{\Gamma}_{k}^{a\dagger}\hat{a}_{k}\right)\,.\end{eqnarray}

Here the rotating-wave and dipole approximations are employed, and
the Hamiltonian terms have the interpretation as follows:

\begin{itemize}
\item $\hat{H}_{1}$ - paraxial mode free Hamiltonian 
\item $\hat{H}_{2}$ - atomic transition free Hamiltonian 
\item $\hat{H}_{3}$ - interferometer and atomic reservoir free Hamiltonians 
\item $\hat{H}_{4}$ - atom-field interaction Hamiltonian 
\item $\hat{H}_{5}$ - atom-reservoir interaction Hamiltonian 
\item $\hat{H}_{6}$ - field-reservoir interaction Hamiltonian 
\end{itemize}
The frequencies $\omega_{k}$ are the mode-frequencies of the $k$-th
interferometer modes, with annihilation operator $\hat{a}_{k}$. The
sum over $k$ is restricted to a single polarization, under the assumption
that only a single polarization of the cavity field is excited here,
with momentum near $\mathbf{k}_{0}$ - which is the longitudinal photon
momentum at the carrier wavelength.

The frequencies $\omega_{\mu}(t)$ are the transition frequencies
of the $\mu$-th atomic transition. In general these may be time-dependent,
for example, if an external magnetic field is used to create a time-varying
Zeeman splitting. The corresponding operators are $\hat{\sigma}_{\mu}^{-}=\left|2\right\rangle _{\mu}\left\langle 1\right|_{\mu}$
and $\hat{\sigma}_{\mu}^{z}=\left|2\right\rangle _{\mu}\left\langle 2\right|_{\mu}-\left|1\right\rangle _{\mu}\left\langle 1\right|_{\mu}$.
Similarly, the coupling term $g_{k}\left(t,\mathbf{r}_{\mu}\right)$
may be time and space-dependent, via the use of a time and space varying
control field. In a pure two-level system, this coupling term would
be expressed as:\begin{equation}
g_{k}\left(t,\mathbf{r}_{\mu}\right)=g\left(t\right)u_{k}\left(\mathbf{r}_{\mu}\right)\end{equation}

Here $g\left(t\right)=\left[\mu^{2}\left(t\right)\omega_{c}/2\hbar\varepsilon_{0}\right]$
, where $\mu\left(t\right)$ is the electric dipole moment of the
atomic transition. This can be made time-dependent in the case of
forbidden transitions in even isotopes of alkaline earths, using a
magnetic control field\cite{Hollberg}. As usual, $u_{k}\left(\mathbf{r}_{\mu}\right)e^{-i\mathbf{k}_{0}\cdot\mathbf{r}_{\mu}}$
is the mode function of a running wave with longitudinal momentum
equal to $\mathbf{k}_{0}$ and a transverse mode structure of $u_{k}\left(\mathbf{r}_{\mu}\right)$,
assumed not to depend on the longitudinal position in the simplest
cases. 

With a three-level atom and electromagnetic control field, the coupling
term has a more complex behaviour that depends on the dynamics of
a third level, which we have assumed can be eliminated if it has a
far-off-resonant Raman coupling. The resulting coupling term has the
structure:\begin{equation}
g_{k}\left(t,\mathbf{r}_{\mu}\right)=g\left(t\right)\Omega\left(\mathbf{r}_{\mu}\right)u_{k}\left(\mathbf{r}_{\mu}\right)\end{equation}

A consequence of this structure of the coupling constant is that there
may be two distinct spatial variations involved: one from the control
field, and one from the stored quantum field. For simplicity, we will
assume a spatially uniform control field intensity so that $\Omega\left(\mathbf{r}_{\mu}\right)=1$
in the following analysis, and we will absorb the phase variation
of the control field into a single mode function $u\left(\mathbf{r}_{\mu}\right)$
with modulus $U_{\mu}$.

Generically, it is possible to divide up the atoms into equivalence
classes with the same coupling constant modulus $U_{j}$ and transition
frequency $\omega_{j}$. If the coupling constant and relevant field
modes have radial symmetry, these correspond to distinct radial shells. 

This creates a set of inequivalent atomic spin operators, defined
as:\begin{eqnarray}
\hat{J}_{j}^{+} & = & \sum_{\mu\in s(j)}\hat{\sigma}_{\mu}^{+}u^{*}\left(\mathbf{r}_{\mu}\right)/U_{j}\nonumber \\
\hat{J}_{j}^{-} & = & \sum_{\mu\in s(j)}\hat{\sigma}_{\mu}^{-}u_{k}\left(\mathbf{r}_{\mu}\right)/U_{j}\nonumber \\
\hat{J}_{j}^{z} & = & \sum_{\mu\in s(j)}\hat{\sigma}_{\mu}^{z}\end{eqnarray}

Initially ignoring (initially) the effects of atomic reservoirs and
losses, which should be small in an atomic system intended for use
as a quantum memory, the resulting Heisenberg picture field and atomic
equations in the rotating-wave and paraxial approximations are as
follows:\begin{eqnarray}
\frac{\partial}{\partial\, t}\hat{a} & = & -\left(i\omega_{0}+\kappa\right)\hat{a}-i\sum_{j}g_{j}\left(t\right)\hat{J}_{{\bf j}}^{-}+\hat{\Gamma}_{k}\nonumber \\
\frac{\partial}{\partial\, t}\hat{J}_{j}^{-} & = & -i\omega_{j}\hat{J}_{j}^{-}+ig_{j}^{*}\left(t\right)\hat{a}\left(t\right)\hat{J}_{{\bf j}}^{z}\nonumber \\
\frac{\partial}{\partial\, t}\hat{J}_{j}^{z} & = & 2\left[ig_{j}\left(t\right)\hat{a}^{\dagger}\hat{J}_{j}^{-}+H.c.\right]\end{eqnarray}
 Here $g_{j}\left(t\right)=g\left(t\right)U_{j}$, and there are also
corresponding equations for conjugate fields. This assumes that the
mode function does not vary rapidly over the location of the grouped
atoms. 

We note here that in general there may be many distinct transverse
electromagnetic mode functions $u_{k}$ that are able to couple to
the atoms. In addition, the cavity loss is at a rate $\kappa_{k}$
due to coupling to the cavity output fields, while $\hat{\Gamma}_{k}$
is the quantum operator for the input and output fields with different
transverse mode indices $k$. According to standard input-output theory\cite{collgard84,gardcoll85},

\begin{equation}
\hat{\Gamma}_{k}=\sqrt{T_{o}/\tau_{r}}\left(\widehat{A}_{k,in}-\widehat{A}_{k,out}\right)\,\,,\end{equation}
 where $\widehat{A}_{k,in}$ is the input photon field and $\widehat{A}_{k,out}$
is the output field, while $T_{o}$ is the mirror transmissivity of
the output coupler, and $\tau_{r}$ is the cavity round-trip time.

In this paper we will only consider the case of a single-mode interferometer
interacting with a non-saturated homogeneous medium, so that $\hat{J}_{j}^{z}\approx-N_{j}$.
We can introduce an effective harmonic oscillator operator of:\begin{equation}
\widehat{b}=\frac{1}{g(t)}\sum_{j}g_{j}\left(t\right)\hat{J}_{{\bf j}}^{-}\,,\end{equation}
 where $g=\sqrt{\sum_{j}N_{j}\left|g_{0,j}^{2}\left(t\right)\right|}$.
We also assume that the medium has a single resonance at $\omega_{j}=\omega$,
which means that there is no inhomogeneous or Doppler broadening.
This would require cooling and possibly trapping in an optical lattice
to eliminate atomic motion. The corresponding Heisenberg equations
are:

\begin{eqnarray}
\frac{\partial}{\partial\, t}\hat{a} & = & -\left[\kappa+i\omega_{0}\right]\hat{a}-ig\left(t\right)\hat{b}+\hat{\Gamma}\nonumber \\
\frac{\partial}{\partial\, t}\widehat{b} & = & -i\omega\hat{b}-ig\left(t\right)\hat{a}_{{\bf k}}\left(t\right)\,.\end{eqnarray}

In a rotating frame resonant with the input carrier frequency of the
quantum signal $\omega_{L}$, this leads to the following effective
Hamiltonian:

\begin{equation}
H=\hbar\delta\widehat{a}^{\dagger}\widehat{a}+\hbar\Delta\widehat{b}^{\dagger}\widehat{b}+\hbar g(t)(\widehat{b}^{\dagger}\widehat{a}+\widehat{a}^{\dagger}\widehat{b})\,.\end{equation}
 where $\delta=\omega_{0}-\omega_{L}$, $\Delta=\omega-\omega_{L}$.
This one-photon detuning $\Delta$ is replaced by the two photon detuning
in the case of a Raman-type interaction.

\subsection{Nanomechanical oscillators}

Similar results are obtained for the effective Hamiltonian of mechanical
oscillators - like an atomic position or nanomechanical oscillator
- in a cavity\cite{Walls,scottkimble,Harris2008}. In this case the
position oscillation has a frequency that is physically analogous
to the separation of the two lower levels in a three-level atomic
model. A control field is needed to create a Raman transition between
the oscillator levels. This type of situation is studied theoretically
as a means of laser cooling nanomechanical oscillators, which has
been recently demonstrated experimentally\cite{Vahala}.

To derive this relationship, we start with a microscopic Hamiltonian
for the radiation field inside an interferometer coupled to a nano-mechanical
oscillator, interacting via the dielectric energy of the coupled system\cite{DrummondQuant}.
This gives a Hamiltonian of form:\begin{equation}
\hat{H}_{nano}=\sum_{j}\hat{H}_{j}\,,\end{equation}
 where the Hamiltonian terms are given by:

\begin{eqnarray}
\hat{H}_{1} & = & \hbar\sum_{k}\omega_{k}\hat{a}_{k}^{\dagger}\hat{a}_{k}\nonumber \\
\hat{H}_{2} & = & \hbar\sum_{\mu}\omega_{j}^{m}\hat{b}_{j}^{\dagger}\hat{b}_{j}\nonumber \\
\hat{H}_{3} & = & \hat{H}^{a}+\hat{H}^{b}\nonumber \\
\hat{H}_{4} & = & \hbar\int d^{3}\mathbf{r}\left(\frac{1}{\varepsilon(\mathbf{r})}-\frac{1}{\varepsilon_{0}}\right)\left|\hat{\mathbf{D}}\left(\mathbf{r}\right)\right|^{2}\nonumber \\
\hat{H}_{5} & = & i\hbar\sum_{j}\left(\hat{\Gamma}_{j}^{b}\hat{b}_{j}^{\dagger}-\hat{\Gamma}_{j}^{b\dagger}\hat{b}_{j}\right)\,\nonumber \\
\hat{H}_{6} & = & i\hbar\sum_{k}\left(\hat{\Gamma}_{k}^{a}\hat{a}_{k}^{\dagger}-\hat{\Gamma}_{k}^{a\dagger}\hat{a}_{k}\right)\,.\end{eqnarray}

Here the Hamiltonian terms have the interpretation:

\begin{itemize}
\item $\hat{H}_{1}$ - paraxial mode free Hamiltonian 
\item $\hat{H}_{2}$ - nano-mechanical oscillator free Hamiltonian 
\item $\hat{H}_{3}$ - interferometer and oscillator reservoir free Hamiltonians 
\item $\hat{H}_{4}$ - interaction energy of the nano-oscillator dielectric
in an external field
\item $\hat{H}_{5}$ - oscillator-reservoir interaction Hamiltonian 
\item $\hat{H}_{6}$ - field-reservoir interaction Hamiltonian 
\end{itemize}
The frequencies $\omega_{k}$ are the mode-frequencies of the $k$-th
interferometer modes, with annihilation operator $\hat{a}_{k}$, as
previously. The frequency $\omega_{j}^{m}$ is the $j-$th resonant
mode frequency of the nano-mechanical oscillator. The field $\hat{\mathbf{D}}\left(\mathbf{r}\right)$is
the electromagnetic displacement field;\begin{equation}
\hat{\mathbf{D}}\left(\mathbf{r}\right)=\sum_{k}\left[\frac{\hbar\omega_{k}\varepsilon\left(\mathbf{r}\right)}{2}\right]\left(u_{k}\left(\mathbf{r}\right)\hat{a}_{k}+H.c.\right)\end{equation}
which is the relevant canonical field variable. We note that for a
standing wave interferometer, with only a single mode of the resonator
and nano-mechanical oscillator, this will reduce to the standard quantum
model of a nano-oscillator as a movable mirror or dielectric inside
a cavity\cite{Walls}:\begin{equation}
H=\hbar\delta\widehat{a}^{\dagger}\widehat{a}+\hbar\omega^{m}\widehat{b}^{\dagger}\widehat{b}+\hbar g\widehat{a}^{\dagger}\widehat{a}(\widehat{b}^{\dagger}+\widehat{b})\,.\end{equation}
Here, $\delta=\omega_{0}-\omega_{L}$, and $\omega^{m}$ is the resonant
frequency of the nanomechanical oscillator.

Since we wish to eliminate the effects of direct radiation pressure
on the oscilator dielectric, we treat a running-wave in which the
field modes by themselves are not coupled to the oscillator motion,
to lowest order. Next, suppose there is an additional counterpropagating
control field $\Omega(t)e^{i\omega_{c}t}$ incident on the oscillator.
This additional field is able to interfere either constructively or
destructively with the intracavity field, at the mirror location.
Let $\omega_{c}=\omega_{0}-\omega$, so the control field is red-detuned
with respect to the Fabry-Perot resonance, which is precisely the
condition required for sideband cooling of a nano-mechanical oscillator.
We will also assume, for simplicity, that the experimental goal of
cooling to the oscillator ground-state is achieved, which means that
the heating rate of the oscillator due to its thermal reservoirs is
sufficiently small.

This leads to the following effective Hamiltonian, in which non-resonant
terms are neglected:

\begin{eqnarray}
H & = & \hbar\delta\widehat{a}^{\dagger}\widehat{a}+\hbar\Delta\widehat{b}^{\dagger}\widehat{b}+\nonumber \\
 & + & \hbar g(\Omega^{*}(t)\widehat{a}\widehat{b}^{\dagger}+\Omega(t)\widehat{a}^{\dagger}\widehat{b})\end{eqnarray}

We see that, for a real control field with $g(t)=g\Omega(t)$, this
expression is identical to the one derived for the case of a weakly
excited atomic resonance.

\subsection{Input/output mode expansions}

As is common in scattering theory, we can define input and output
modes corresponding to two distinct Hilbert spaces for the asymptotic
past and future of the memory. We limit ourselves to treating a single
transverse mode $A_{0}^{in}(t)$ for simplicity. A complete mode expansion
into longitudinal modes of the incoming external field for past times
$t<0$ is\begin{equation}
\widehat{A}_{0}^{in}(t)=\sum_{n}\widehat{a}_{n}^{in}u_{n}^{in}(t)\,,\end{equation}
 where $\widehat{A}_{0}^{in}$ is a boson input field such that $\left[\widehat{A}_{0}^{in}(t),\widehat{A}_{0}^{in\dagger}(t')\right]=\delta(t-t')$.
Here the $a_{n}^{in}$ are bosonic mode operators {and $u_{n}^{in}(t)$
the mode functions, }whose expectation values determine the incoming
pulse shape. Similarly, the operator $\widehat{A}_{0}^{out}(t)$ is
the quantum operator for the output field. A complete mode expansion
for the outgoing external field after a memory storage time $T$ is
an expansion over future times ($t>T$):\begin{equation}
\widehat{A}_{0}^{out}(t)=\sum_{n}\widehat{a}_{n}^{out}u_{n}^{in}(t)\,,\end{equation}
 where the $\widehat{a}_{n}^{out}$are also boson annihilation operators,
{and the $u_{n}^{out}(t)$ the output mode functions}. We focus
on the simplest possible case of single longitudinal mode storage
devices, which are designed to accurately write into memory, store
then read out information for one input and one output bosonic mode.
The single-mode input and output operators of the states to be {}``remembered''
will be labelled $\widehat{a}_{n}^{in}$ and $\widehat{a}_{n}^{out}$
. To simply the typography, we will omit the caret on single mode
operators $a$, $b$ and fields $A$, $B$ in the remaining sections.

\begin{figure}
\includegraphics[width=0.8\columnwidth]{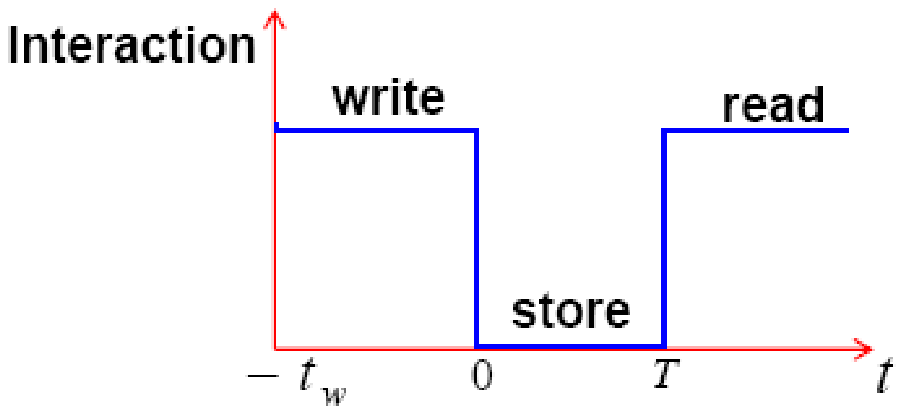}

\caption{Memory involves three stages: writing, reading and storing. The interaction
is turned on, then off, then on, in a controllable way. \label{fig:Memory-involves-three}}

\end{figure}

\section{Memory Fidelity}

It is crucial to determine the level of memory performance and accuracy
at which one can convincingly claim a {}``quantum memory''. A standard
figure of merit for memory performance is that of the average fidelity
$\bar{F}$ between input and output states, as defined over a predetermined
set of input states. Here the output state is a density matrix $\hat{\rho}_{out}$,
which is obtained on tracing the output state over the input modes
and loss reservoirs:\begin{eqnarray}
\hat{\rho}_{out} & = & Tr_{r}\left[|\Psi_{out}\rangle\langle\Psi_{out}|\right]\nonumber \\
 & = & Tr_{r}\left[\hat{U}|\Psi_{in}\rangle\langle\Psi_{in}|\hat{U}^{-1}\right]\,.\end{eqnarray}

We will be considering pure state inputs, in which case the average
fidelity is defined as:

\begin{equation}
\bar{F}=\int P(\Psi_{in})\langle\Psi_{in}|\hat{\rho}_{out}(\Psi_{in})|\Psi_{in}\rangle d\mu(\Psi_{in})\end{equation}
 Here $P(\Psi_{in})$ is the probability of using a given state $\Psi_{in}$,
while $\hat{\rho}_{out}(\Psi_{in})$ is the output density matrix
conditioned on input of $\Psi_{in}$, and $d\mu(\Psi_{in})$ is the
integration measure used over the set of input states.

The average fidelity obtained must be compared with the best average
fidelity possible using a `classical' measure, store and prepare strategy,
in order to claim that one has a quantum memory. There is no known
limit to which quantum states may be feasibly prepared, nor on what
observables can be measured, except that the commutators of quantum
mechanics prevent simultaneous, precise measurement of non-commuting
variables. This means that the set of inputs used is important in
establishing fidelity bounds. For example, if the input states are
orthogonal - like the number states - then the classical fidelity
bound is unity. All the number states can in principle be measured
using a perfect photo-detector, the corresponding number recorded
and stored, followed by regeneration of the original number state
with perfect fidelity.

This means that superpositions must be an integral part of the input
alphabet of quantum states. An important issue is that the relative
phase of superpositions must be recalled in a quantum memory device.
Thus, the fidelity can not be measured in the same way as the photon
counting efficiency: a memory that generates outputs with random phases
will have a high photon-counting efficiency, but a low quantum fidelity.
This is because the fidelity measure is phase-sensitive, which is
essential for a quantum memory. To experimentally characterize a quantum
memory it is therefore necessary to measure input and output states
interferometrically. Measuring the energy efficiency alone cannot
rule out memory phase errors caused, for example, by timing jitter
in the control signals.

\subsection{Linear memory}

In this paper, we treat linear memory models, with all reservoirs
in the vacuum state, and with no excess phase noise. This type of
memory has the useful property that it is able, ideally, to preserve
any input state with a subsequent time-delayed read-out. 

In quantum mechanics, a given initial state $|\Psi_{in}\rangle$ in
the Schroedinger picture is transformed to a final state by making
a unitary transformation on the input Hilbert space:\begin{equation}
|\Psi_{out}\rangle=\hat{U}|\Psi_{in}\rangle\,.\end{equation}
In greater detail, we can divide the Hilbert space into the input
space, output space, and reservoir space consisting of all other degrees
of freedom. We assume that initially the input space has a factorized
state :

\begin{equation}
|\Psi_{in}\rangle=|\psi_{0}\rangle_{in}|0\rangle_{out}|0\rangle_{r}\end{equation}
The purpose of a quantum memory is to transform this input state into
an output state at a later time, with the structure:\begin{equation}
|\Psi_{out}\rangle=|0\rangle_{in}|\psi_{0}\rangle_{out}|0\rangle_{r}\end{equation}

It is convenient to describe the input in terms of a function of input
mode creation operators $a_{0}^{\dagger}$ defined at $t=-\infty$,
so that:\begin{equation}
|\psi_{0}\rangle=f\left(a_{0}^{\dagger}\right)|0\rangle_{in}\end{equation}

We will find in the next sections that in the Heisenberg picture,
the overall effect of either losses or mode mis-matching is identical
to a (time-delayed) beam-splitter with transmission efficiency $\eta_{M}$,
so that the memory output state is:\begin{equation}
|\Psi_{out}\rangle=|0\rangle_{in}f\left(a_{0}^{\dagger}(\infty)\right)|0\rangle_{out}|0\rangle_{r}\end{equation}
where:\begin{eqnarray}
a_{0}(\infty) & = & \sqrt{\eta_{M}}a_{0}+\sqrt{1-\eta_{M}}a_{0}^{r}\,.\label{eq:beam-splitter}\end{eqnarray}
 Here $a_{0}$ is now understood to act on the output vacuum state,
and $a_{0}^{r}$ is a bosonic operator which only acts on the zero-temperature
reservoir, so that $\langle a_{0}^{r\dagger}a_{0}^{r}\rangle_{r}=0$.

Ideal performance is obtained when retrieval efficiency $\eta_{M}=1$,
so that the input and output mode operators are identical, apart from
the technical issue that they are defined on different Hilbert spaces.
In practice, loss and noise will be introduced at all three stages
of a quantum memory: not all information can be retrieved, since $\sqrt{\eta_{M}}<1$.

\subsection{Coherent state memories}

The most common set of input states considered to date are coherent
states, which have already proved useful to quantum applications such
as teleportation\cite{furteleport} and quantum state transfer from
light onto atoms\cite{Julsgaard}. If we consider our input set as
the set of coherent states with a Gaussian distribution $P(\alpha)=1/(\overline{n}\pi)e^{-|\alpha|^{2}/\overline{n}}$,
and mean photon number $\overline{n}$, the fidelity average measure
$\overline{F}$ is \begin{equation}
\bar{F}_{\bar{n}}^{g}=\int P(\alpha)\langle\alpha|\hat{\rho}_{out}(\alpha)|\alpha\rangle d^{2}\alpha\,,\label{eq:fpa}\end{equation}
 where $\hat{\rho}_{out}(\alpha)$ is the output state for the coherent
input state $|\alpha\rangle$.

The results of Hammerer et al\cite{Hammerer} and Braunstein et al\cite{braun}
show that for any \emph{classical} channel, the average fidelity is
constrained by \begin{equation}
\bar{F}_{\bar{n}}^{g}\leq(1+\overline{n})/(2\overline{n}+1)\,.\label{eq:f}\end{equation}
 Thus, the result $\bar{F}_{\bar{n}}^{g}>(1+\overline{n})/(2\overline{n}+1)$
serves as a benchmark for the claim of a quantum memory of coherent
states.

\begin{figure}[H]
\includegraphics[width=0.95\columnwidth]{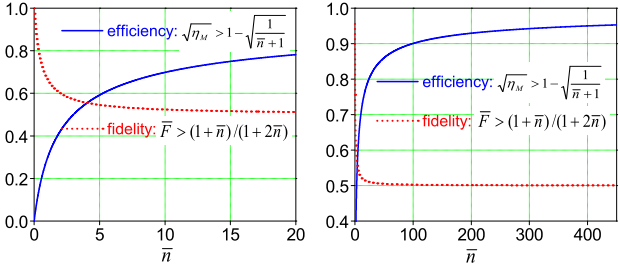}

\caption{Fidelity $\bar{F}_{\bar{n}}^{c}$ (dashed) and corresponding beam
splitter efficiency $\sqrt{\eta_{M}}$ (solid) required for quantum
memory as a function of $\bar{n}$ for coherent input states. High
$\bar{n}$ will require a high efficiency $\eta_{M}$, for the claim
of a quantum memory, whereas for low $\bar{n}$ , quantum memory is
achievable for lower efficiencies. Horizontal lines indicate the respective
classical bounds. \label{fig:Fidelity and efficiency }}

\end{figure}

We calculate $\bar{F}_{\bar{n}}^{c}$ for our beam-splitter solution
Eq. (\ref{eq:beam-splitter}). In this solution, the output is $\hat{\rho}_{out}(\alpha)=|\sqrt{\eta_{M}}\alpha\rangle\langle\sqrt{\eta_{M}}\alpha|$.
Simple calculation gives \begin{equation}
\bar{F}_{\bar{n}}^{c}=\frac{1}{1+\overline{n}(1-\sqrt{\eta_{M}})^{2}}\,.\label{eq:n}\end{equation}
 The condition for quantum memory (so that (\ref{eq:f}) is violated)
is thus satisfied for efficiencies

\begin{equation}
\sqrt{\eta_{M}}>1-\sqrt{\frac{1}{\overline{n}+1}}\,.\label{eq:nm}\end{equation}
 We note that for $\overline{n}\geq20,$ the bound follows an almost
flat line relation to $\bar{n}$, which is the well known flat distribution
for which fidelity $\bar{F}_{\infty}^{c}>0.5$ is required for a quantum
memory\cite{Hammerer,Julsgaard,furteleport}. These fidelities correspond
in the beam splitter memory to quite high efficiencies, so for $\overline{n}=20$,
quantum memory is achieved for $\sqrt{\eta_{M}}>0.78$. For $\overline{n}$
small, say $\overline{n}=1$, which requires fidelity $\bar{F}_{1}^{c}>2/3,$
we note that quite low efficiencies ($\sqrt{\eta_{M}}>0.293$) are
enough for a claim of a quantum memory (Fig. \ref{fig:Fidelity and efficiency }).

\subsection{Arbitrary state memories}

An ideal quantum memory must do more than just store coherent states.
For many quantum information applications, the quantum states that
must be stored may be in a larger class of possible quantum inputs.
Recent experiments and theory have investigated other possibilities,
like squeezed states\cite{sqbenchmark,Appel,honda}. The most general
case is a completely arbitrary quantum input state. However, it is
essential to bound the input energy in some way. Otherwise, the averages
are dominated by inputs of infinitely large energy, that no physical
memory could possibly store without giving rise to a black hole.

Here we define the input state as any possible state with a maximum
photon number less than $n_{m}$. This corresponds to an arbitrary
state $\left|\vec{\Psi}\right\rangle $ of $n_{m}$ levels, where:\begin{equation}
\left|\vec{\Psi^{in}}\right\rangle =\sum_{n=0}^{n_{m}-1}\Psi_{n}\left|n\right\rangle \,.\end{equation}
 so that the highest photon number is $n=n_{m}-1$. The fidelity average
$\overline{F}_{n_{m}}$ is then the average fidelity over all possible
coefficients $\vec{\Psi}$, satisfying the constraint that $\left|\vec{\Psi}\right|=1$
, i.e: \begin{equation}
\bar{F}_{n_{m}}=\frac{\int\delta(\left|\vec{\Psi}\right|-1)\langle\vec{\Psi}|\hat{\rho}_{out}(\vec{\Psi})|\vec{\Psi}\rangle d^{2n_{m}}\vec{\Psi}}{\int\delta(\left|\vec{\Psi}\right|-1)d^{2n_{m}}\vec{\Psi}}\,,\label{eq:fpa2}\end{equation}
 where $\hat{\rho}_{out}(\vec{\Psi})$ is the output reduced density
matrix for the arbitrary bounded input state $|\vec{\Psi}\rangle$,
after tracing over any reservoirs coupled to the memory.

To determine the classical fidelity limit in this case, we recall
that there is a known fidelity limit for (imperfect) cloning of an
arbitrary $n_{m}$ level state, to produce an infinitely large number
of copies. This limit is that \cite{classical clone}:\begin{equation}
\bar{F}_{n_{m}}\le\frac{2}{n_{m}+1}\,.\end{equation}
 Since a classical memory can clearly generate any number of copies
of a quantum state, this result shows that for any \emph{classical}
memory with an arbitrary input of bounded maximum photon number, the
average fidelity is constrained by the one-to-many cloning limit.

\begin{figure}[H]
\includegraphics[width=0.9\columnwidth]{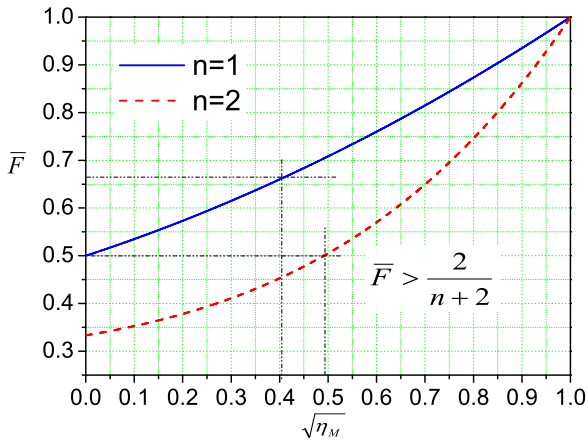}

\caption{Average fidelity vs beam splitter efficiency of a quantum memory for
arbitrary input states with up to $n=1$ (solid line) and $n=2$ (dashed
line) photons.\label{fig:Average-fidelity-vs}}

\end{figure}

We now calculate $\bar{F}_{n_{m}}$ for our beam-splitter solution
Eq. (\ref{eq:beam-splitter}). The total input state, including a
reservoir labelled $r$ and assumed to be a vacuum state, is:\begin{equation}
\left|\Psi_{T}^{in}\right\rangle =\sum_{n=0}^{n_{m}-1}\frac{\Psi_{n}}{\sqrt{n!}}\hat{a}^{\dagger n}|\mathbf{0}\rangle\,.\end{equation}
 Here $\Psi_{n}$ is the probability amplitude for the $|n\rangle$
input state. The output state is therefore:\begin{eqnarray}
\left|\Psi^{out}\right\rangle  & = & \hat{U}\left|\Psi_{T}^{in}\right\rangle \nonumber \\
 & = & \sum_{n=0}^{n_{m}-1}\frac{\Psi_{n}}{\sqrt{n!}}\left[\hat{a}^{out\dagger}\right]^{n}|\mathbf{0}\rangle\,\\
 & = & \sum_{n=0}^{n_{m}-1}\frac{\Psi_{n}}{\sqrt{n!}}\left[\sqrt{\eta_{M}}a_{0}^{in\dagger}+\sqrt{1-\eta_{M}}a_{0}^{r\dagger}\right]^{n}|\mathbf{0}\rangle\,.\nonumber \end{eqnarray}
 We can now calculate the fidelity in the case of $n_{m}=2$ and $n_{m}=3$,
which allows for arbitrary states with up to $1$ and $2$ photons
respectively. Since the reservoir modes are not the input to the memory,
we trace over the mode $r$, to obtain the predicted memory fidelities
\begin{eqnarray}
\overline{F}_{2} & = & \frac{\eta_{M}+2\sqrt{\eta_{M}}+3}{6}\nonumber \\
\overline{F}_{3} & = & \frac{\eta_{M}^{2}+2\eta_{M}\sqrt{\eta_{M}}+3\eta_{M}+2\sqrt{\eta_{M}}+4}{12}\,.\end{eqnarray}
 for 2 and 3-dimensional (up to $1$ and $2$ photon number) input
states respectively. These results are graphed below, in Fig. \ref{fig:Average-fidelity-vs}.

It is straightforward to prove, using SU(n) symmetry, that in the
limit of zero efficiency the quantum memory will have an average fidelity
of $\overline{F}_{n_{m}}=1/n_{m}$ . This is always less than the
fidelity achievable by a classical {}``measure and regenerate''
strategy. In general, the best classical average fidelity decreases
as the number of possible quantum levels increases. This is easily
understandable: a single measurement gives very little information
about the coherent superpositions that may exist in a quantum state
with many levels. For this reason, an arbitrary quantum state fidelity
measure gives a much better indication of the power of a quantum memory
than a measure constrained to a single set of states like the coherent
states. This gives a strong motivation for more general experimental
tests of quantum memory performance.

\section{Q-switched memory dynamics: mode matching}

In the previous sections, we calculated the fidelity where the relation
between the input and output states is describable by the beam splitter
solution Eq. (\ref{eq:beam-splitter}). Now, we show under which conditions
this solution is predicted. To understand the role of mode-matching,
we examine in this section the simple model of an empty Q-switched
cavity.

We consider first a simplistic quantum memory model of an empty Q-switched
cavity, tuned to frequency $\omega_{0}=\omega_{L}+\delta$. In practice,
long storage times are not readily achievable without a separate oscillator
such as an atom medium for storage. However, we analyse this model
first to develop an understanding of the dynamics of the three stages
of memory process: writing, storage and reading. The corresponding
effective internal Hamiltonian is:\begin{equation}
\widehat{H}=\hbar\delta a^{\dagger}a\,.\end{equation}

The cavity is partially transmitting, with variable cavity decay rate
$\kappa(t)$, allowing a coupling between the cavity mode $a$ and
a pulsed input field $a_{in}(t)$. For a cavity whose only loss is
through one mirror acting as an input/output coupler, the dynamical
Heisenberg equation linking input and cavity mode operators is\cite{collgard84,gardcoll85}
\begin{equation}
\dot{a}=-\left[i\delta(t)+\kappa(t)\right]a+\sqrt{2\kappa(t)}A^{in}(t)\,.\label{eq:cavitydynamiceqn}\end{equation}
 The writing stage begins at $-t_{w}$ (Fig. \ref{fig:Memory-involves-three})
and is of duration up to $t=0$. Defining a time-evolution function:
\begin{equation}
T_{\kappa}(t,t')=\exp\left[-\int_{t'}^{t}\left[i\delta(\tau)+\kappa(\tau)\right]d\tau\right]\,,\end{equation}
 the interaction given by Eq. (\ref{eq:cavitydynamiceqn}) has the
general solution \begin{eqnarray}
a(t) & = & T_{\kappa}(t,-t_{w})a(-t_{w})+\nonumber \\
 & + & \int_{\tau=-t_{w}}^{t}T_{\kappa}(t,\tau)\sqrt{2\kappa(\tau)}A^{in}(\tau)d\tau\,.\end{eqnarray}

The purpose of the memory is to read in the field at $t<0$, and then
output selected information after a memory time $T$. We therefore
introduce a model decay rate with Q-switching between a large value
$\kappa$ and a small value $\kappa_{S}$, at zero detuning:\begin{eqnarray}
\kappa(t) & = & \kappa\,\,\,[t<0]\nonumber \\
\kappa(t) & = & \kappa_{S}\,\,\,[0<t<T]\nonumber \\
\kappa(t) & = & \kappa\,\,\,[t>T]\,.\end{eqnarray}
 We note here as a practical issue that all cavities have excess loss
and noise over and above that given just by considering input/output
couplers. This may be unimportant during the input/output stages,
when $\kappa(t)$ is large. However, it is certainly significant when
$\kappa(t)$ is small. For this reason, $\kappa_{S}$ and the corresponding
vacuum reservoir term must include all losses during the storage time,
including loss in the dielectric coatings and diffraction losses.
Additional phase-noise and corresponding phase-relaxation terms due
to acoustic noise are ignored for simplicity.

We note our model quantum memory has a time-reversal symmetry around
$t=T/2$, since $\kappa(t)=\kappa(T-t)$. This is not essential, since
one could easily choose $\kappa(t>T)\neq\kappa(t<0)$ . However, this
feature - which is also found in some other memory proposals - provides
a useful insight into design of a quantum memory, and the mode-functions
that are coupled into and out of the memory. Here, of course, time-reversal
implies reversing the propagation direction of all fields, including
the input and output fields. A typical input-output relation with
some residual loss during the storage time is shown in Fig. {\ref{fig:Empty-cavity}.
This is obtained from a numerical solution of Eq. (\ref{eq:cavitydynamiceqn})
in a P-representation\cite{Prep}, which transforms the operator equations
into c-number equations. In this case, the input state of the field
is assumed to be a coherent state. The calculated solution clearly
displays the time-reversal. We note that the calculation can be extended
to an arbitrary initial state using the positive P-representation
method\cite{posP}.

\begin{figure}
\includegraphics[width=0.9\columnwidth]{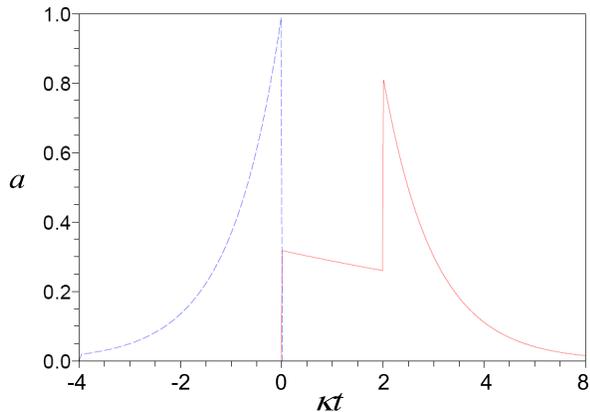}

\caption{Q-switched cavity input (dashed blue line) and output (solid red line)
amplitudes with $\kappa=1$, $\kappa_{S}=0.1$, $T=2.0$. Input mode
shape is mode matched to the time-reversed cavity decay.\label{fig:Empty-cavity}}

\end{figure}

To explain the operation of the Q-switched quantum memory more clearly,
we seek analytical solutions, and now expand the incoming and outgoing
field operators into past-time ($t<0$) and future time ($t>T$) modes.
This allows us to easily distinguish what is stored in the memory
in the past from what is read out, in the future.

\subsection{Writing: past-time modes}

Our model gives for the stored cavity mode solution, when the cavity
coupling is switched to a small value for storage: \begin{equation}
a(0)=a_{0}^{in}\equiv\sqrt{2\kappa}\int_{-\infty}^{0}e^{\kappa\tau}A^{in}(\tau)d\tau\,.\end{equation}
 Where we have considered $\delta=0$ for simplicity, which means
that the cavity is resonant with the field carrier frequency. We have
allowed the writing time $t_{w}$ to be infinite, in practice of duration
much longer than pulse durations and cavity lifetimes, so as to erase
information associated with the initial cavity solution.

We note the operator $A^{in}(t)$ is the quantum operator for the
input field, but the coupling to the cavity is such that only a certain
mode of this incoming field is effectively coupled. We choose our
input mode expansion to be: \begin{equation}
u_{n}(t)=\sqrt{2\kappa}e^{\kappa t}L_{n}(-2\kappa t)\Theta(-t)\,,\end{equation}
 which are modified Laguerre polynomials. Since the Laguerre polynomials
are a complete set, any incoming waveform that vanishes as $t\rightarrow-\infty$
can be represented as a linear combination of Laguerre functions.
Introducing $z=-2\kappa t$, these have orthogonality relations of:\begin{eqnarray}
\int_{-\infty}^{0}u_{n}^{in}(t)u_{m}^{in*}(t)dt & = & \int_{0}^{\infty}e^{-2z}L_{n}(z)L_{m}(z)dz\nonumber \\
 & = & \delta_{mn}\,.\end{eqnarray}
 In this expansion, $u_{n}(t)$ are orthogonal mode functions on the
space of past times, prior to switching on the memory at $t=0$, so
that:\begin{equation}
a_{n}^{in}=\int_{-\infty}^{0}A^{in}(t)u_{n}^{in*}(t)dt\,.\end{equation}
 Thus, using the field commutators we obtain the following bosonic
commutators for $a^{in}$: \begin{eqnarray}
\left[a_{n}^{in},a_{m}^{in\dagger}\right] & = & \int_{-\infty}^{0}\int_{-\infty}^{0}u_{n}^{in*}(t)u_{m}^{in}(t')\delta(t-t')dtdt'\nonumber \\
 & = & \int_{-\infty}^{0}u_{n}^{in*}(t)u_{m}^{in}(t)dt\nonumber \\
 & = & \delta_{nm}\,.\end{eqnarray}
 Due to the orthogonality of the Laguerre functions only the $u_{0}(t)$
term will give a nonzero contribution to $a(0)$. To gain maximum
efficiency of the memory, the experimentalist must therefore construct
the incoming pulse shape to match this mode, so that $<a_{n}^{in}>=\delta_{0n}$.
With this choice, when evaluating expectation values we can effectively
simplify to a single input mode: \begin{equation}
A^{in}(t)=u_{0}^{in}(t)a_{0}^{in}=\sqrt{2\kappa}e^{\kappa t}a_{0}^{in}\,.\label{eq:awithmode}\end{equation}

We note that this saw-tooth type mode structure is time-asymmetric
(see Fig. \ref{fig:Empty-cavity}), which is not ideal in terms of
mode matching to the typical Gaussian pulses produced by mode-locked
lasers. Improved matching to symmetric pulses could be realized through
more careful shaping of the cavity coupling in time, ie, making $\kappa(t)$
a prescribed shape.

We also stress that this cavity-based memory is a strictly mono-mode
memory, from a temporal point of view. One temporal mode only is stored,
the others being reflected. No bipartite (or n-partite) states composed
of two or more temporal modes $u_{i}(t)$ can thus be stored. This
device can however be used as a mode-converter to manipulate temporal
multi-mode quantum states.

\subsection{Storage period}

In the simplest model, in which no medium is present and cavity losses
are assumed zero, the value $a(0)$ is stored with maximum efficiency
in the cavity for a duration $T$, so that \begin{equation}
a(T)=a(0)\,.\end{equation}
 More generally, there is a residual storage loss $\kappa_{S}$ at
this stage. The dynamical Eq. (\ref{eq:cavitydynamiceqn}) applies
again, but this time as we have no pulsed input, the input $A^{in}(\tau)$
represents only the incoming vacuum field. To make a clear distinction
between the two inputs, we will denote $A^{in}(t)\equiv A_{v}^{in}(t)$
where $t>T$, so that:

\begin{equation}
a(T)=e^{-\kappa_{S}T}a(0)+\sqrt{2\kappa_{S}}\int_{\tau=o}^{T}e^{\kappa_{S}(\tau-T)}A_{v}^{in}(\tau)d\tau\,.\end{equation}

When there are excess losses in addition to output coupler loss, $A_{v}^{in}(\tau)$
must include all the relevant loss reservoirs associated with $\kappa_{S}$
. Although we do not consider this in detail, there can also be additional
noise sources which will degrade the stored quantum information. These
include thermal noise if the signal is at relatively low frequency,
as in microwave experiments, and additional phase noise from acoustic
noise or $1/f$ noise in the mirrors and dielectrics. Phase-noise
can become very significant in the limit of long storage times, and
must be considered when storage fidelity is measured.

\subsection{Reading: future-time modes}

At time $T$, the output stage commences, and the cavity is switched
back to a large $\kappa$, to allow transmission, or reading, of the
remembered signal outside the cavity. The solution is \begin{equation}
a(t)=e^{-\kappa(t-T)}a(T)+\sqrt{2\kappa}\int_{T}^{t}e^{\kappa(\tau-t)}A_{v}^{in}(\tau)d\tau\,.\end{equation}
 We focus on the output field transmitted through the cavity, given
by\cite{collgard84,gardcoll85} \begin{equation}
A^{out}=\sqrt{2\kappa}a-A_{v}^{in}\,.\end{equation}

Making use of the time-reversal symmetry of our model, we will choose
the output modes to be the time-reversed input modes, so that \begin{eqnarray}
u_{n}^{out}(t) & = & u_{n}^{in*}(T-t)\nonumber \\
 & = & \sqrt{2\kappa}e^{-\kappa(t-T)}L_{n}(2\kappa(t-T))\,.\end{eqnarray}
 Introducing $z=2\kappa(t-T)$, these have orthogonality relations
in future time, of:\begin{eqnarray}
\int_{T}^{\infty}u_{n}^{out}(t)u_{m}^{out*}(t)dt & = & \int_{0}^{\infty}e^{-z}L_{n}(z)L_{m}(z)dz\nonumber \\
 & = & \delta_{mn}\,.\end{eqnarray}
 At this point, we note that maximum efficiency of retrieval is achieved
if we temporally match the output with the input in the following
way. We define the filtered output field operator as:\begin{eqnarray}
a_{0}^{out} & \equiv & \int_{T}^{\infty}u_{0}^{*}(t)A^{out}(t)dt\,\nonumber \\
 & = & \sqrt{2\kappa}\int_{T}^{\infty}e^{-\kappa(t-T)}A^{out}(t)dt\,.\end{eqnarray}
 We find\begin{eqnarray}
a_{0}^{out} & = & \sqrt{2\kappa}\int_{T}^{\infty}e^{-\kappa(t-T)}A^{out}(t)dt\nonumber \\
 & = & \sqrt{2\kappa}\int_{T}^{\infty}e^{-\kappa(t-T)}(\sqrt{2\kappa}a(t)-A_{v}^{in}(t))dt\nonumber \\
 & = & 2\kappa\int_{T}^{\infty}e^{-2\kappa(t-T)}a(T)dt\nonumber \\
 &  & -2\kappa\sqrt{2\kappa}\int_{T}^{\infty}e^{-\kappa(t-T)}dt\int_{T}^{t}e^{\kappa(\tau-t)}A_{v}^{in}(\tau)d\tau\nonumber \\
 &  & +\sqrt{2\kappa}\int_{T}^{\infty}e^{-\kappa(t-T)}A_{v}^{in}(t)dt\nonumber \\
 & = & a(T)\,.\end{eqnarray}
 In the ideal case with $\kappa_{S}=0$, we know that $a(T)=a(0)=a_{0}^{in}$,
so we retrieve the signal $a_{0}^{in}$, while all information related
to unwanted vacuum inputs at future times, $A_{v}^{in}$, is completely
absent from the filtered output. The explanation of this desirable
behaviour is rather simple. After $t=T$, the cavity is perfectly
matched as an absorber of \emph{incoming} vacuum modes to the future-time
$u_{0}$ mode. As a result, the cavity now absorbs all the incoming
vacuum field radiation in the incoming $n=0$ future-time mode, while
simultaneously emitting the stored information in an outgoing $n=0$
future-time mode. In summary, while the modes with $n>0$ are simply
reflected, the stored $n=0$ mode changes places with an incoming
$n=0$ vacuum mode.

Thus, an incoming past-time $n=0$ mode is time-delayed by the memory
time $T$, then re-emitted into an outgoing future-time $n=0$ mode.
This is readable without losses (in the ideal case) using a temporal
mode filter. We note that the pulse-shape of the output mode is time-reversed
with respect to the input mode.

In our model of an empty Q-switched cavity with perfect temporal mode-matching
and loss occurring during storage, the storage cannot be ideal. The
presence of losses means not all information can be retrieved due
to the residual loss $\kappa_{S}$ from the cavity over the storage
time of duration $T$. This means that $a(T)\neq a(0)$. Instead\begin{eqnarray}
a(T) & = & e^{-\kappa_{S}T}a(0)+\sqrt{2\kappa_{0}}\int_{0}^{T}e^{\kappa_{S}(t-T)}A_{v}^{in}(t)dt\nonumber \\
 & = & \sqrt{\eta_{M}}a_{0}^{in}+\sqrt{1-\eta_{M}}a_{v}^{in}\,,\end{eqnarray}
 where the overall memory efficiency is given by:\begin{equation}
\sqrt{\eta_{M}}=e^{-\kappa_{S}T}\,.\end{equation}

\section{Storage using a linear atomic medium}

Since all cavities leak or absorb photons, information from the input
field is better stored using long-lived atomic transitions. In some
experiments, a control field is used to determine whether a particular
atomic transition can decay, to release photons into the cavity mode.
With the control field off, emission of the quanta is suppressed.
We thus propose a simple model in which the cavity decay is now fixed
at $\kappa$. The interaction of the cavity field with the linear
medium is switched on, to write, then off, to store, and finally on
again, to allow readout of the stored quantum information.

At a fixed detuning, the coupling between the cavity field and the
medium is modelled by the interaction Hamiltonian \begin{equation}
H=\hbar\delta a^{\dagger}a+\hbar\Delta b^{\dagger}b+\hbar g(t)(b^{\dagger}a+a^{\dagger}b)\,.\label{eq:control field shaping}\end{equation}

This model may describe, for example, a three-level Raman experiment
operated near resonance with detuning $\Delta$, in the linear response
regime without saturation. Here the coupling $g(t)$ is modulated
with a control field at a different wavelength to the signal field.

Alternatively, one may wish to consider experiments where the effective
coupling is switching using time-varying detunings $\delta(t)$, $\Delta(t)$:
\begin{equation}
H=\hbar\delta(t)a^{\dagger}a+\hbar\Delta(t)b^{\dagger}b+\hbar g(b^{\dagger}a+a^{\dagger}b)\,.\label{eq:time-varying detunings}\end{equation}
 This scenario is found in experiments which employ Zeeman, Stark
or two-photon control field shifting to change detunings. This strategy
can be used in a range of experiments from solid-state crystals and
cold atoms to artificial-atom experiments using superconducting cavities
and transmission lines.

\subsection{Input (writing): }

During the input stage, the interaction is switched on. We assume
for simplicity that all couplings and detunings are held constant
and that $\delta=0$, so that the Heisenberg evolution equations of
the system operators are:

\begin{eqnarray}
\frac{da(t)}{dt} & = & -\kappa a(t)-igb(t)+\sqrt{2\kappa}A^{in}(t)\label{eq:set}\\
\frac{db(t)}{dt} & = & -(\gamma+i\Delta)b(t)-iga(t)+\sqrt{2\gamma}B_{v}^{in}(t)\,,\nonumber \end{eqnarray}
 where $\gamma$ is the atomic decay rate. In these equations the
source term proportional to $B_{v}^{in}(t)$ corresponds to the coupling
of the medium with their respective baths, whereas for $a(t)$ the
input field corresponds to the incoming field we wish to store. These
equations are valid both for two-level atoms interacting with one
field in an optical cavity and for three-level atoms in a Raman configuration
when the excited level can be adiabatically eliminated.

To solve the system of equations, it is useful to rewrite as \begin{eqnarray}
\frac{d}{dt}\vec{\alpha} & =- & \mathbf{G}\vec{\alpha}+\vec{\alpha}^{in}\,,\label{eq:lineatom}\end{eqnarray}
 where $\vec{\alpha}=\left(\begin{array}{c}
a\\
b\end{array}\right)$, $\vec{\alpha^{in}}=\left(\begin{array}{c}
\sqrt{2\kappa}A^{in}\\
\sqrt{2\gamma}B_{v}^{in}\end{array}\right)$ and{ \begin{eqnarray}
\mathbf{G} & = & \left(\begin{array}{cc}
\kappa & ig\\
ig & \gamma+i\Delta\end{array}\right)\nonumber \\
 & = & \frac{\kappa-\gamma-i\Delta}{2}\sigma_{z}+ig\sigma_{x}+\frac{\kappa+\gamma+i\Delta}{2}\nonumber \\
 & = & \kappa_{-}\sigma_{z}+ig\sigma_{x}+\kappa_{+}\,.\end{eqnarray}
 } Here we have defined $\kappa_{\pm}=\left[\kappa\pm(\gamma+i\Delta)\right]/2$
and introduced the Pauli spin matrices.

Defining a time-evolution matrix using a time-ordered exponential
as \begin{equation}
\mathbf{T}_{G}(t,t')=T:\left\{ \exp\left[-\int_{t'}^{t}\mathbf{G}(\tau)d\tau\right]\,\right\} :\,,\end{equation}
 the operator solution of Eq. (\ref{eq:lineatom}) is\begin{eqnarray}
\vec{\alpha}(t) & = & e^{-\mathbf{G}(t-t_{0})}\alpha(t_{0})+\int_{-t_{w}}^{t}e^{-\mathbf{G}(t-\tau)}\vec{\alpha}^{in}d\tau\,.\end{eqnarray}
 In the limit of interest where the writing time, starting at $t=-t_{w}$,
is long and we stop writing at $t=0$, the initial cavity operators
decay, and the solution becomes\begin{equation}
\vec{\alpha}(0)=\int_{-\infty}^{0}e^{\mathbf{G}\tau}\vec{\alpha}^{in}d\tau\,.\end{equation}
 Simplifying, we note that we can re-express this using: \begin{equation}
e^{\mathbf{G}\tau}=e^{\kappa_{+}\tau}e^{\overrightarrow{m}\cdot\overrightarrow{\sigma}\tau}\,,\end{equation}
 where \begin{equation}
\overrightarrow{m}=(ig,0,\kappa_{-})\,,\end{equation}
 and: \begin{equation}
\vec{\sigma}=(\sigma_{x},\sigma_{y},\sigma_{z})\,.\end{equation}
 Since an exponentiated sum of Pauli matrices can be expanded in elementary
form using: \begin{equation}
e^{\overrightarrow{m}\overrightarrow{\sigma}\tau}=ch(m\tau)\mathbf{I}+\frac{\overrightarrow{m}.\overrightarrow{\sigma}}{m}sh(m\tau)\,,\end{equation}
 where $\mathbf{I}$ is the $2\times2$ identity matrix, we abbreviate
$ch\equiv cosh$ and $sh\equiv sinh$, and take $m=\sqrt{\kappa_{-}^{2}-g^{2}}$.
We have $\overrightarrow{m}.\overrightarrow{\sigma}=\left(\begin{array}{cc}
\kappa_{-} & ig\\
ig & -\kappa_{-}\end{array}\right)$. Thus we find the general solution for the input process:\begin{eqnarray}
\vec{\alpha}(t) & = & \int_{-\infty}^{0}e^{\kappa_{+}\tau}[ch(m\tau)+\,\nonumber \\
 & + & \frac{1}{m}sh(m\tau)\left(\begin{array}{cc}
\kappa_{-} & ig\\
ig & -\kappa_{-}\end{array}\right)]\vec{\alpha}^{in}(\tau)d\tau\,.\end{eqnarray}

Our final stored solutions are written

\begin{eqnarray}
a(0) & = & \sqrt{2\kappa}\int_{-\infty}^{0}e^{\kappa_{+}\tau}[ch(m\tau)+\frac{\kappa_{-}sh(m\tau)}{2m}]A^{in}(\tau)d\tau\nonumber \\
 &  & +\sqrt{2\gamma}\int_{-\infty}^{0}e^{\kappa_{+}\tau}[\frac{igsh(m\tau)}{m}]B_{v}^{in}(\tau)d\tau\,,\end{eqnarray}

\begin{eqnarray}
b(0) & = & \sqrt{2\kappa}\int_{-\infty}^{0}e^{\kappa_{+}\tau}\frac{ig}{m}sh(m\tau)A^{in}(\tau)d\tau\nonumber \\
 &  & +\sqrt{2\gamma}\int_{-\infty}^{0}e^{\kappa_{+}\tau}[ch(m\tau)-\frac{\kappa_{-}sh(m\tau)}{2m}]B_{v}^{in}(\tau)d\tau\nonumber \\
 & = & \sqrt{2\kappa}\int_{-\infty}^{0}e^{\kappa_{+}\tau}\frac{ig}{m}sh(m\tau)a_{0}^{in}u_{0}^{in}(\tau)d\tau+\mathcal{B}\,,\end{eqnarray}
 where $\mathcal{B}$ represents all the additional noise terms, dependent
on $B_{v}^{in}$. We express $A^{in}$ in terms of the input mode
function $u_{0}^{in}(\tau$), as in (\ref{eq:awithmode}). The $b(0)$
represents the stored mode of the signal $A^{in}(\tau)$. This result
implies an optimal choice of pulse shape for $u_{0}^{in}(\tau)$,
to maximise memory efficiency. In particular, we will choose\begin{equation}
u_{0}^{in}(t)=\sqrt{2(\kappa\gamma+g^{2})(\kappa+\gamma)}e^{\kappa_{+}^{*}\tau/2}\frac{-i}{m}sh(m\tau)\,.\end{equation}
 In contrast to the Q-switched cavity memory memory, the typical duration
of the pulse mode giving the higher transfer efficiency is not merely
$1/\kappa$ (i.e.: the inverse of the cavity bandwidth). Here, the
duration of the adapted pulse depends strongly on the relative values
of the cavity coupling rate $\kappa$ and the atom-light coupling
rate $g$. In practice, a pulse as short as possible is preferable
to prevent relaxation. Accordingly, a critically damped regime corresponding
to $m=0$ should be chosen if possible.

\subsection{Storage: }

We store the recorded state in the medium for a time $T$. Here the
control field is off, and there is no interaction between the cavity
and medium, so that $g=0$. Similar results are found if we assume
that $\Delta$ is very large, which also suppresses the coupling between
atoms and cavity. A real non-ideal memory will have nonzero atomic
and cavity loss $\gamma$ and $\kappa$. The solutions at the end
of the storage time are then:\begin{eqnarray}
a(T) & = & a(0)e^{-\kappa T}+\sqrt{2\kappa}\int_{0}^{T}e^{-\kappa(T-t)}A_{v}^{in}(t)dt\,,\nonumber \\
b(T) & = & b(0)e^{-(\gamma+i\Delta)T}+\nonumber \\
 &  & +\sqrt{2\gamma}\int_{0}^{T}e^{-(\gamma+i\Delta)(T-t)}A_{v}^{in}(t)dt\,.\end{eqnarray}

\subsection{Output (reading)}

After a time $T$, the control field is switched on, but with only
the vacuum input to the cavity, and the medium coupled to the cavity
mode. The cavity end-mirror has finite transmission, so the signal
can be read outside the cavity. Reading is a dynamical process for
times $t>T$, described by (\ref{eq:lineatom}), to give intracavity
solutions\begin{eqnarray}
\overrightarrow{\alpha}(t) & = & e^{-\mathbf{G}(t-T)}\overrightarrow{\alpha}(T)+\nonumber \\
 &  & +\int_{T}^{t}e^{-\mathbf{G}(t-\tau)}\vec{\alpha}_{v}^{in}(\tau)d\tau\,.\end{eqnarray}

The solution for the cavity field $a(t)$ is therefore: \begin{eqnarray}
a(t) & = & e^{-\kappa_{+}(t-T)}[ch(m(t-T))a(T)-\nonumber \\
 &  & \frac{sh(m(t-T))}{m}\{a(T)\kappa_{-}+igb(T)\}]\nonumber \\
 &  & +\int_{T}^{t}e^{-\kappa_{+}(t-\tau)}\{\sqrt{2\kappa}[ch(m(t-\tau))\nonumber \\
 &  & -\frac{\kappa_{-}}{m}sh(m(t-\tau))]A_{v}^{in}(\tau)\nonumber \\
 &  & -\sqrt{2\gamma}[\frac{ig}{m}sh(m(t-\tau))]B_{v}^{in}(\tau)\}d\tau\,.\end{eqnarray}
We also have for the field output\cite{collgard84,gardcoll85} \begin{equation}
A^{out}(t)=\sqrt{2\kappa}a(t)-A_{v}^{in}(t)\,.\end{equation}

\section{Comparison of memory strategies}

We will now compare in detail two possible strategies for gating the
quantum memory: a fixed detuning method with variable coupling, and
a fixed coupling method with variable detuning. Thus, we analyse in
turn the outputs for two models of Eq. (\ref{eq:control field shaping})
and Eq. (\ref{eq:time-varying detunings}), where the coupling between
the cavity field and the medium is switched by $g(t)$ or a time-varying
detuning $\Delta(t)$ respectively.

\subsection{Fixed detuning ($\Delta=0$)}

\begin{figure}
\includegraphics[width=1\columnwidth]{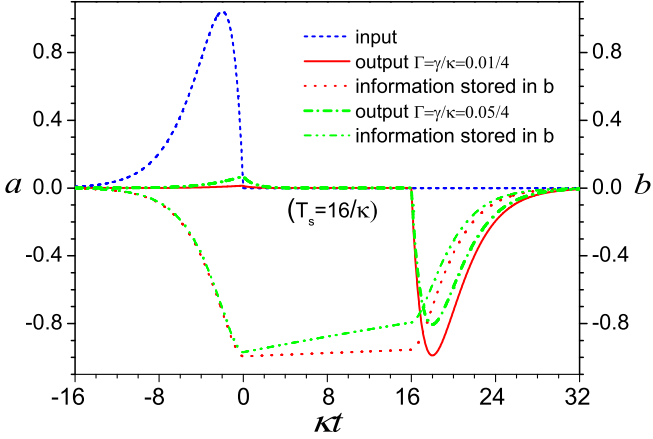}

\caption{{Atomic-coupled cavity input (dashed blue line), and output amplitudes
with $\gamma/\kappa=0.01/4$ (solid black line), $\gamma/\kappa=0.05/4$
(dash-dotted green line) for the zero detuning strategy $\Delta=0$.
Corresponding dashed thin lines represent the information $b(t)$
stored in the atom. Here $C\simeq100$ and $T_{s}=16/\kappa$, and
the critically damped case applies. }\textcolor{red}{\label{fig:different gamma}}}

\end{figure}

The coupling $g(t)$ is given as

\begin{eqnarray}
g(t) & = & g\,\quad[t<0]\nonumber \\
g(t) & = & 0_{}\,\,\,[0<t<T]\nonumber \\
g(t) & = & g\,\quad[t>T]\,.\end{eqnarray}
 Using $\kappa_{\pm}=(\kappa\pm\gamma)/2$ due to $\Delta=0$, we
obtain the relation between the operators $a(0),$ $b(0)$ and $a_{0}^{in}$:\begin{eqnarray}
a(0) & = & \sqrt{2\kappa}\sqrt{2(\kappa\gamma+g^{2})(\kappa+\gamma)}\int_{-\infty}^{0}e^{2\kappa_{+}\tau}\frac{-i}{m}\nonumber \\
 &  & \times[ch(m(\tau)+\frac{\kappa_{-}sh(m\tau)}{2m}]sh(m\tau)a_{0}^{in}d\tau+noise\nonumber \\
 & = & \frac{\sqrt{\kappa}\gamma i}{\sqrt{(\kappa\gamma+g^{2})(\kappa+\gamma)}}a_{0}^{in}+noise\,,\end{eqnarray}

\begin{eqnarray}
b(0) & = & \sqrt{2\kappa}\sqrt{2(\kappa\gamma+g^{2})(\kappa+\gamma)}\frac{g}{m^{2}}\nonumber \\
 &  & \times\int e^{2\kappa_{+}\tau}sh^{2}(m\tau)a_{0}^{in}d\tau+noise\nonumber \\
 & = & \frac{\sqrt{\kappa}g}{\sqrt{(\kappa\gamma+g^{2})(\kappa+\gamma)}}a_{0}^{in}+noise\,.\end{eqnarray}

\begin{figure}
\includegraphics[width=0.9\columnwidth]{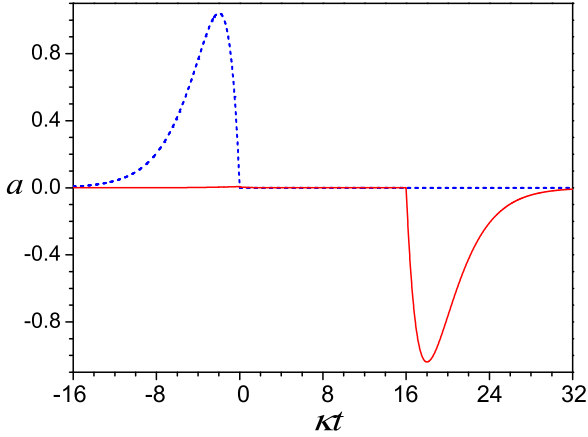}

\caption{Atomic-coupled cavity input (dashed blue line) and output (solid red
line) amplitudes in the zero detuning $\Delta=0$ case with $\kappa=4$,
$g=2$, $\gamma=0.01$, $T_{S}=4.0$, using direct numerical integration.
Input mode shape is mode matched to the critical cavity decay.\label{fig:atomic-coupled cavity}}

\end{figure}

After a time $T$, the time reversed $g(t)$ retrieves the cavity
mode into the output mode $u_{0}^{out*}(t)=u_{0}^{in}(T-t)$, which
is the time reverse of $u_{0}^{in}(t)$. The optimal function for
the cavity output pulse is thus\begin{eqnarray}
a_{0}^{out} & = & \int_{T}^{\infty}u_{0}^{out*}(t)a_{out}(t)dt\,,\nonumber \\
a_{out}(t) & = & \sqrt{2\kappa}a(t)-A_{v}^{in}(t)\,.\end{eqnarray}

After calculating the relevant integrals, but omitting the explicit
form of the {}``noise'' terms, we have

\begin{eqnarray}
a_{0}^{out} & = & \frac{\sqrt{\kappa}\gamma ia(T)+\sqrt{\kappa}gb(T)}{\sqrt{(\kappa\gamma+g^{2})(\kappa+\gamma)}}+noise\nonumber \\
 & = & \frac{-\kappa\gamma^{2}e^{-\kappa T}+\kappa g^{2}e^{-(\gamma+i\Delta)T}}{(\kappa\gamma+g^{2})(\kappa+\gamma)}a_{0}^{in}+noise\nonumber \\
 & = & \sqrt{\eta_{M}}a_{0}^{in}+\sqrt{1-\eta_{M}}a_{0}^{r}\end{eqnarray}
 which reduces to (\ref{eq:beam-splitter}) where $a_{0}^{r}$ is
the reservoir mode arising from the {}``noise'' term, and $\sqrt{\eta_{M}}$
is the overall memory efficiency given by

\begin{eqnarray}
\sqrt{\eta_{M}} & = & \frac{\kappa g^{2}e^{-\gamma T}-\kappa\gamma^{2}e^{-\kappa T}}{(\kappa\gamma+g^{2})(\kappa+\gamma)}\nonumber \\
 & = & \frac{Ce^{-\gamma T}}{(1+C)(1+\Gamma)}-\frac{\Gamma e^{-\kappa T}}{(1+C)(1+\Gamma)}.\label{eq:efficiency}\end{eqnarray}
 Here, we introduce the cooperativity parameter $C=g^{2}/\kappa\gamma$
and $\Gamma=\gamma/\kappa$.{ This result agrees with that obtained
previously \cite{theory cavity}, in the limit of $C\gg\Gamma$, or
$\kappa T$ large enough so that the second term is negligible. The
optimal case is to ensure large $C\gg\Gamma$, $C\gg1$, large $\kappa$
compared to $\gamma$, so $\Gamma$ is small. It is still necessary
however to ensure that the storage time is small enough so that $\gamma T\ll1$.
However, $T$ can be many cavity lifetimes, $\kappa T\gg1$. We note
we do not want $\Gamma=1$ because critical damping would require
zero $g$. If $m=0$ , so that $g=\kappa_{-}=(\kappa-\gamma)/2$,
we obtain the critically damped case for which the desired input temporal
mode function is}

\begin{equation}
u_{0}^{in}(t)=\frac{-i\kappa_{+}}{\sqrt{2}}\sqrt{\kappa_{+}}e^{\kappa_{+}^{}t/2}t\,.\label{eq:mode function}\end{equation}

Fig. \ref{fig:different gamma} shows the typical input-output relation
for various loss ratios during the storage time of duration $T$.
For the same cavity damping $\kappa$, different rates of optical
coherence decay will result in different memory efficiencies. For
$\gamma=0.01,$ $\sqrt{\eta_{M}}=0.95$, while for $\gamma=0.05,$
$\sqrt{\eta_{M}}=0.80.$ We can use the ratio of the integral of envelope
between $a_{0}^{out}$ and $a_{0}^{in}$ $[\int u_{0}^{out*}(t)a_{out}(t)dt/\int_{-\infty}^{0}u_{0}^{in*}(t)a_{in}(t)dt$]
to check the value of $\sqrt{\eta_{M}}$. If $\gamma$ is larger,
the atomic lifetime is shorter, which means the information stored
in the medium decays more quickly (shown by thin dashed green curve),
resulting in a reduced efficiency.

In summary, with an appropriate selection of mode-matched filters,
we are still able to retrieve the input signal with high efficiency,
provided $\Gamma\ll1$. {The results are confirmed by numerical integration
of the coupled cavity-oscillator equations, as shown in Fig. \ref{fig:atomic-coupled cavity}.
This numerical method thus serves as a way to explore more sophisticated
nonlinear models of the atomic medium.}%
\begin{figure}
\includegraphics[width=0.9\columnwidth]{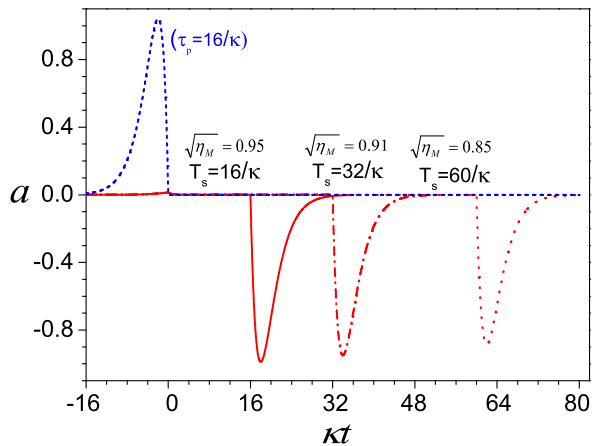}

\caption{Atomic-coupled cavity input (dashed blue line) and {output (solid
black line) amplitudes with $\gamma/\kappa=0.01/4$ and different
storage time $T_{S}=4$, 8, 15. }Input mode shape is mode matched
to the critical cavity decay{. $\kappa=4$, $g=2$.} \textcolor{red}{\label{fig:long memory time}}}

\end{figure}

To analyse the effectiveness of the memory as a quantum memory, we
must calculate the mean fidelity. Here we consider, for definiteness,
the simplest encoding strategy with coherent states. Other strategies
- for example using an arbitrary state with photon number bounds -
will generally have different thresholds, as explained in Section
(II).

For the case of $\overline{n}=1$, any retrieval with $\sqrt{\eta_{M}}>0.293$
can be claimed to be a {}``quantum memory''. For $\overline{n}>20$
(Fig. \ref{fig:Fidelity and efficiency }), the curve of required
average fidelity as a function of mean number of photons is very flat
and close to the classical boundary \cite{nature05136}, which is
why low photon numbers are preferable in experiments on quantum memory,
if high fidelity is required. At $\overline{n}=20$ we need a much
higher retrieval efficiency of $\sqrt{\eta_{M}}>0.80$ to ensure the
device is a true quantum memory.

A long storage time $T$ is consistent with high memory fidelity $\bar{F}$
(Fig. \ref{fig:long memory time}), provided we optimise for high
efficiency using mode matching, and provided the atomic losses are
not significant over the storage time ($\gamma T\ll1$). For an input
signal duration $\tau_{p}=4$, with residual loss $\gamma=0.01$,
we get a retrieval efficiency $\sqrt{\eta_{M}}=0.95$, $0.91$, $0.85$
for the storage times $T=4$, $8$, $15$ respectively. The average
fidelities are $\overline{F}=0.95,$ $0.86$, $0.69$, respectively,
all of them larger than the classical bound $\overline{F}=0.51$ required
for a quantum memory at $\overline{n}=20$. Thus, for these parameters,
with input states giving $\overline{n}=20$, we are able to predict
the existence of a quantum memory, with both high fidelity and relatively
long memory lifetime.{ At lower photon numbers of $\overline{n}\sim1$,
a much higher loss is possible before loss of quantum memory.}

\subsection{Time-varying detuning}

In experiments using two-level atoms one may control the coupling
by with a time-varying detuning $\Delta(t)$ \cite{hetet}. During
writing and reading the atoms are strongly coupled to the field to
allow transfer of the quantum state. During storage, the coupling
is decreased by using a greatly increased detuning, controllable via
a magnetic field or a Stark shift. To model this case, we employ a
time-varying detuning with $\Delta_{L}\gg\kappa,$$\gamma$:}

{\begin{equation}
\begin{array}{ccccc}
\Delta(t) & = &  & 0 & [t<0]\\
\Delta(t) & = &  & \Delta_{L} & [0<t<T/2]\\
\Delta(t) & = & - & \Delta_{L} & [T/2<t<T]\\
\Delta(t) & = &  & 0 & [t>T]\,.\end{array}\end{equation}
 Here, the storage period is divided into two parts with opposite
detunings in order to ensure the phase is the same between signal
and output field. In the writing and reading periods, choosing critical
damping $g=\kappa_{-}=(\kappa-\gamma)/2$ expressed in real terms
with detuning $\Delta=0,$ we will have the same input mode $u_{0}^{in}(t)$
as above. The overall memory efficiency in this case is}

\begin{eqnarray}
\sqrt{\eta_{M}} & = & \frac{4\kappa(g^{2}e^{-\gamma T}-\gamma^{2}e_{}^{-\kappa T})}{(\kappa+\gamma)^{3}}\,,\end{eqnarray}
 which is the same form as Eq. (\ref{eq:efficiency}) for the critically
damped case.

\begin{figure}[h]
\includegraphics[width=0.9\columnwidth]{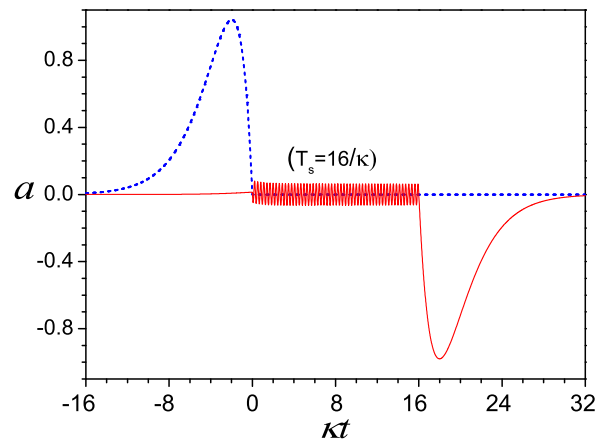}

\caption{{Atomic-coupled cavity input (dashed blue line) and output amplitudes
with $\gamma/\kappa=0.01/4$ (solid red line). }Input mode shape
is mode matched to the critical cavity decay{. $\kappa=4$, $g=(\kappa-\gamma)/2$,
$T_{S}=4.0$,$\Delta_{L}=27\pi$.} \textcolor{red}{\label{fig:time-varying detuning}}}

\end{figure}

The atomic-coupled cavity input and output amplitudes with $\gamma=0.01,$
$\Delta_{L}=27\pi$ is shown in Fig. \ref{fig:time-varying detuning}.
The dashed black line represents the desired output mode shape matched
to the critical cavity decay.

\section{Summary}

We consider a general protocol for a dynamical quantum memory, using
a cavity-oscillator model. Our definition of an acceptable quantum
memory is based on two elementary criteria. To qualify as a quantum
device, it must have a fidelity over a given set of input states that
is better than any classical measure and regenerate strategy. To qualify
as a memory it must be able store the input state over a time-scale
longer than the input signal duration.

We analyse fidelity measures using both a coherent state input and
an arbitrary quantum superposition input. Our general conclusion is
that an optimal memory performance of a quantum memory is obtained
through mode-matching the input pulse shape to a specific input mode
of the memory device.

Three models of quantum memory are considered, of increasing complexity.
All the models possess a time-reversal symmetry, so that output modes
are obtained through a time-reversal of the input modes. First, to
introduce the importance of temporally mode-matching the input pulse
to the cavity mode, we consider a simple Q-switched cavity. This is
sensitive to cavity losses during the storage period, which are difficult
to eliminate.

Next, we introduce a model of a linearly coupled atomic memory, including
losses, but with step-function modulation of the coupling. Provided
a suitably modified asymmetric temporal mode is used, the effects
of cavity loss are suppressed for long atomic lifetimes, and it is
possible to largely decouple the input quantum mode from the lossy
intracavity field mode. We show that there is an optimal coupling
strength which generates a mode-matched input and output pulse. Finally,
we consider a model in which the detuning is modulated in time, and
show that this has a similar behaviour to the modulated coupling protocol.

With tailored input and output mode shapes, this type of quantum memory
device promises to give both relatively long memory lifetimes and
high memory quality.

\begin{acknowledgments}
We thank the Australian Research Council for support through ARC Centre
of Excellence and Discovery grants. Ecole Normale Superieure and Universite
Pierre et Marie Curie also provided support through their visiting
professor programs. We are grateful to S. Parkins, K. Lehnert, Ping
Koy Lam and others for stimulating discussions. 
\end{acknowledgments}

\end{document}